\documentclass{article}

\PassOptionsToPackage{numbers}{natbib}
\usepackage[final]{neurips_2019}
\usepackage{caption}

\usepackage[utf8]{inputenc} % allow utf-8 input
\usepackage[T1]{fontenc}    % use 8-bit T1 fonts
\usepackage{hyperref}       % hyperlinks
\usepackage{url}            % simple URL typesetting
\usepackage{booktabs}       % professional-quality tables
\usepackage{amsfonts}       % blackboard math symbols
\usepackage{nicefrac}       % compact symbols for 1/2, etc.
\usepackage{microtype}      % microtypography
\usepackage{physics}
\usepackage{graphicx}
\usepackage{dsfont}
\usepackage{xcolor}
\usepackage{textcomp}

\title{Advances in Quantum Deep Learning: An Overview}
 
\author{
Siddhant Garg\thanks{\ \ \ Equal contribution by authors}
\qquad
Goutham Ramakrishnan\footnotemark[1]\\
Department of Computer Sciences\\
University of Wisconsin--Madison\\
\texttt{\{sidgarg,gouthamr\}@cs.wisc.edu}
}

\begin{document}

\maketitle

\begin{abstract}
The last few decades have seen significant breakthroughs in the fields of deep learning and quantum computing. 
Research at the junction of the two fields has garnered an increasing amount of interest, which has led to the development of quantum deep learning and quantum-inspired deep learning techniques in recent times. 
In this work, we present an overview of advances in the intersection of quantum computing and deep learning by discussing the technical contributions, strengths and similarities of various research works in this domain. 
To this end, we review and summarise the different schemes proposed to model quantum neural networks (QNNs) and other variants like quantum convolutional networks (QCNNs). 
We also briefly describe the recent progress in quantum inspired classic deep learning algorithms and their applications to natural language processing.

\end{abstract}

\section{Introduction}
\label{sec:intro}

In recent years, deep neural networks have led to breakthroughs in several domains of machine learning, such as computer vision \cite{7298965, He_2016_CVPR,7298594}, natural language processing \cite{NIPS2017_7181, devlin-etal-2019-bert}, reinforcement learning \cite{silver2016go}, speech recognition \cite{deng2013speech}, etc. 
Deep Learning~\cite{lecun2015deeplearning} forms the backbone of a majority of modern machine learning techniques and has become one of the most active research areas in computer science, spurred on by increased availability of data and computational resources.

Parallelly, there has been remarkable progress in the domain of quantum computing focused towards solving classically intractable problems through computationally cheaper techniques. A major leap forward in quantum computing came when Shor~\cite{10.1137/S0097539795293172,10.1109/SFCS.1994.365700} proposed his famous algorithm for prime factoring numbers in polynomial time, which exposed the vulnerabilities of security protocols such as RSA. 
Consequent research has been aimed at developing poly-time alternatives of classical algorithms utilising the core idea of quantum superposition and entanglement. 
We briefly describe these ideas when reviewing basic principles of quantum computing.

Quantum computing naturally lends its ideas to the domain of machine learning and consequently there been active research on trying to use principles of quantum computing to improve the representation power and computational efficiency of classical ML approaches. 
Quantum extensions to classical ML problems have gained prominence in recent times, such as clustering~\cite{lloyd2013quantum, NIPS2019_8667, otterbach2017unsupervised},
support vector machines~\cite{Rebentrost_2014}, 
gradient descent for linear systems~\cite{Kerenidis_2020},
principal component analysis~\cite{Lloyd_2014},
Boltzmann machines~\cite{Amin_2018},
feature extraction~\cite{wilson2018quantum},
recommendation systems~\cite{kerenidis2016quantum}, 
EM algorithm for Gaussian Mixture Models~\cite{kerenidis2019quantum}, 
variational generations for adversarial learning~\cite{romero2019variational}, etc. 

The perceptron~\cite{rosenblatt1958perceptron} represents a single neuron and forms the basic unit of the deep learning architectures. 
The idea of a quantum perceptron was first proposed by \citet{10.1016/0020-0255(94)00095-S} in 1995 and has since been formalized in multiple works~\cite{GUPTA2001355, NIPS2003_2363, journals/qip/SchuldSP14, Wan_2017,cao2017quantum,Daskin_2018, Farhi2018ClassificationWQ, shao2018quantum,Beer2020quantum}. 
Recently, \citet{wiebe2014quantum} showed that quantum computing can provide a more comprehensive framework for deep learning than classical computing and can help optimization of the underlying objective function.

In this work, we summarise the different ideas presented in the domain of quantum deep learning which include quantum analogues to classic deep learning networks and quantum inspired classic deep learning algorithms. 
We present the different schemes proposed to model quantum neural networks (QNNs) and their corresponding variants like quantum convolutional networks (QCNNs). 

This work is structured as follows: we first review the basics of classical deep learning and quantum computing in Sections~\ref{sec:classical_dl} and \ref{sec:quantum_basics}, for the benefit of an uninitiated reader. 
In Section~\ref{sec:qnn}, we provide a detailed overview of Quantum Neural Networks as formulated in several works, by examining its individual components analogous to a classical NN. 
We also briefly summarize several variants of QNNs and their proposed practical implementations. 
In Section~\ref{sec:quantum_cnn_rnn}, we review works that develop quantum analogues to classical convolutional and recurrent neural networks (CNNs and RNNs). 
In Section~\ref{sec:applications}, we mention several classical deep learning algorithms which have been inspired by quantum methods, including applications to natural language processing. 

\section{Basic Principles of Classical Deep Learning}
\label{sec:classical_dl}
Neural networks represent a subset of machine learning methods, which try to mimic the structure of the human brain in order to learn. 
\textit{Neurons} are the fundamental computational units in neural networks. 
Each neuron performs a sequence of computations on the inputs it receives to give an output. 
Most commonly, this computation is a linear combination of the inputs followed by a non-linear operation, i.e. the output is $F(\sum_{i=1}^{N} w_i x_i)$ where $x_i$ are the inputs to the neuron. 
The $w_i$ are the parameters of the neuron, and $F(.)$ is the non-linear function. 
Commonly used non-linear functions are the Rectified Linear Unit (ReLU) and the sigmoid function $\sigma$: 
$$
ReLU(x) = max(0, x) \quad , \quad \sigma(x) = \frac{1}{1+e^{-x}}
$$

\begin{figure}
\centering
\begin{minipage}{.41\textwidth}
  \centering
  \includegraphics[width=1.8in]{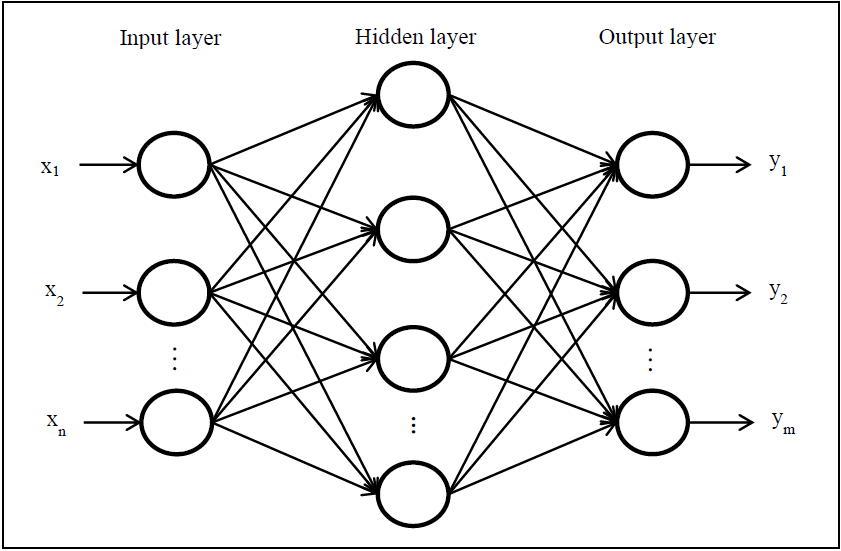}
  \captionof{figure}{A simple feedforward NN with one hidden layer (Figure from \cite{ahmadian2015})}
  \label{fig:simple_nn}
\end{minipage}%
\hspace{10pt}
\begin{minipage}{.55\textwidth}
  \centering
  \includegraphics[width=3in]{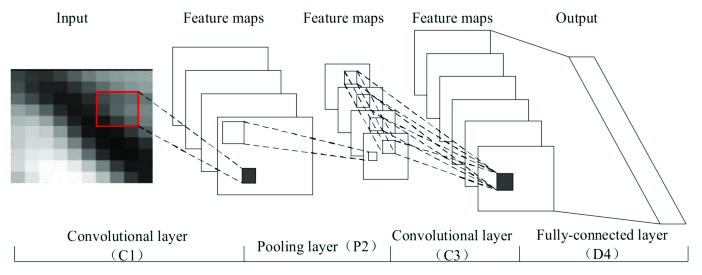}
  \captionof{figure}{The structure of a simple convolutional neural network (Figure from \cite{zhu2018})}
  \label{fig:cnn}
\end{minipage}
\end{figure}

Neural network architectures are constructed by stacking neurons. In fully-connected feedforward neural networks, the output of each neuron in the previous layer is fed to each neuron in the next layer. The simplest neural network is the fully-connected network with one hidden layer (Figure \ref{fig:simple_nn}). Let the input $x$ be $d_1$-dimensional and output be $d_2$-dimensional. Then, a NN a single hidden layer of $h$ units performs the following computation:
$$
N(x) = W_2 \cdot F(W_1 \cdot x)
$$
$W_1$ and $W_2$ are weight matrices of dimensions $h \cross d_1$ and $d_2 \cross h$ respectively. The non-linear function $F$ is applied element-wise to the vector input. This can be generalized to a NN with $M$ hidden layers as: 
$$
N(x) = W_{M+1} \cdot F(W_{M} \cdot F(W_{M-1} \cdot ... \ F(W_1 \cdot x)...)
$$

The universal approximation theorem \cite{citeulike:3561150, journals/nn/LeshnoLPS93} states that, a neural network with a single hidden layer can approximate any function, under assumptions on its continuity. 
However, it is known that deeper networks (with greater number of hidden layers) learn more efficiently and generalize better than shallow networks \cite{NIPS2014_5422, AAAI1714849}. 
Increased availability of data, greater complexity of tasks and the development of hardware resources such as GPUs 
% and TPUs \cite{JouYou17} 
have led to the use of deeper and deeper neural networks, thus the term `deep learning'.

% \begin{figure}
%     \centering
%     \includegraphics[width=2in]{images/simple_nn.png}
%     \caption{The structure of a simple feedforward NN with one hidden layer (Figure from \cite{ahmadian2015})}
%     \label{fig:simple_nn}
% \end{figure}

\paragraph{Parameter Learning}
Like most machine learning algorithms, tasks in deep learning are posed as \textit{Empirical Risk Minimization} (ERM) problems. 
Fundamentally, the parameter learning is done through gradient based optimization methods to minimize a \textit{loss function}. 
The loss function is computed over the training data, and depends on the task at hand. 
Common loss functions include the 0/1 and cross-entropy loss for classification tasks, $l_2$-loss for regression tasks and reconstruction loss for autoencoder tasks (a form of unsupervised learning). 
The backpropagation algorithm \cite{Rumelhart:1986we,lecun1989cnn} uses the chain-rule to offer a computationally efficient way of obtaining gradients in neural networks.
Learning is known to be highly sensitive to the optimization algorithm \cite{kingma2014method,lequoc2011opti} as well as the initialization of the parameters \cite{glorot2010}.

\paragraph{Complex Neural Architectures}
The past decades of deep learning research have led to several breakthroughs such as convolutional neural networks \cite{fukushima:neocognitronbc,Lecun2000,KriSut12Imagenet} (designed for learning hierarchical and translation-invariant features in images), recurrent neural networks \cite{10.5555/104279.104293,hochreiter1997long} (for sequential data such as time series and natural language), ResNets \cite{he2015deep} (designed to combat the vanishing gradient problem in deep learning) and Transformers \cite{vaswani2017attention} (the current state of the art method in natural language processing).  

% \begin{figure}
%     \centering
%     \includegraphics[width=3in]{images/cnn.png}
%     \caption{The structure of a simple CNN (Figure from \cite{zhu2018})}
%     \label{fig:cnn}
% \end{figure}

\paragraph{CNNs}
Convolutional neural networks (CNNs) have revolutionized the field of computer vision, since \citet{lecun1989cnn} demonstrated how to use back propagation to efficiently learn feature maps. They form the basis of most state-of-the-art tasks in modern computer vision, and are widely deployed in applications including image processing, facial recognition, self-driving cars, etc. 

Classical CNNs are designed to capture hierarchical learning of translation-invariant features in structured image data, through the use of convolutional and pooling layers.
Convolutional layers consist of multiple convolutional filters, each of which computes an output feature map by convolving local subsections of the input iteratively. 
Pooling layers perform subsampling to reduce the dimensionality of the feature maps obtained from convolutional layers, most commonly by taking the maximum or mean of several nearby input values. 
A non-linear activation is usually applied to the output of the pooling layer. 

A typical CNN architecture for image classification consists of several successive blocks of convolutional$\rightarrow$pooling$\rightarrow$non-linear layers, followed by a fully connected layer (Figure \ref{fig:cnn}). 
Convolutional filters learn different input patterns, at different levels of abstraction depending upon the depth of the layer. 
For image inputs, the initial layers of the CNN learn to recognize simple features such as edges. 
The features learnt by successive layers become increasingly complex and domain specific, through a combination of features learnt in previous layers. 
CNNs are a powerful technique, and several papers have adapted its ideas to the quantum setting, and we discuss these in Section \ref{sec:quantum_cnn_rnn}.

\begin{figure}
    \centering
    \includegraphics[width=3in]{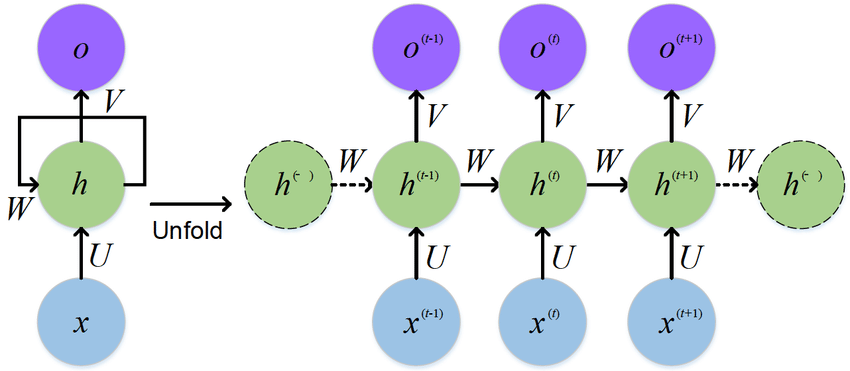}
    \caption{The structure and temporal unfolding of a simple RNN (Figure from \cite{feng2017})}
    \label{fig:rnn}
\end{figure}

\paragraph{RNNs}
Feedforward neural networks are constrained as they perform predefined computations on fixed-size inputs. 
Recurrent Neural Networks (RNNs) are designed to handle sequences of inputs, operating on one input at a time while retaining information about preceding inputs through a hidden state. 
For a sequential input $x=(x^{(1)},\dots,x^{(L)})$, the simplest RNN performs the following computation:
$$
h_t = F(x_t, h_{t-1}), \ \ o_t=G(h_t),  \ \ t=1,\dots,L
$$
$h_t$ and $o_t$ refer to the hidden state and output of the RNN at step $t$ of the sequence, $h_0$ is the initial hidden state, $F$ and $G$ are functions to be learnt. 
RNNs can also be used to learn representations of sequence inputs for different down stream tasks with the final hidden state $h_L$ as the \textit{embedding} of the input $x$. 
Figure \ref{fig:rnn} shows the the temporal unfolding of a simple RNN. 

RNNs are trained using Backpropagation-through-time (BPPT) \cite{58337}, a temporal extension of the backpropagation algorithm.
The versatility of RNNs is such that they are used for a wide variety of applications: sequential-input single-output (e.g. text classification), single-input sequential-output (e.g. image captioning) and sequential-input sequential-output (e.g. part-of-speech tagging, machine translation) tasks.  
Several innovations have improved the performance of the vanilla RNN described above, such as LSTM \cite{hochreiter1997long} and GRU \cite{chung2014}, bidirectional RNNs \cite{650093}, attention mechanism \cite{bahdanau2014neural}, encoder-decoder architecture \cite{cho2014} and more. 

\section{Principles of Quantum Computing}
\label{sec:quantum_basics}

The \textit{qubit} is the basic unit of information in quantum computing. 
The power of quantum computing over classical computing derives from the phenomena of \textit{superposition} and \textit{entanglement} exhibited by qubits. 
Unlike a classical bit which has a value of either 0 or 1, superposition allows for a qubit to exist in a combination of the two states. 
In general, a qubit is represented as:
$$
\ket{\psi} = \alpha \ket{0} + \beta \ket{1}
$$
$\ket{0}$ and $\ket{1}$ represent the two computational \textit{basis} states, $\alpha$ and $\beta$ are complex amplitudes corresponding to each, satisfying $|\alpha|^2 + |\beta|^2 = 1$. 
\textit{Observing} a qubit causes a collapse into one of the basis states. 
The probability of each state being observed is proportional to the square of the amplitude of its coefficient, i.e. the probabilities of observing $\ket{0}$ and $\ket{1}$ are $|\alpha|^2$ and $|\beta|^2$ respectively. 
A qubit is physically realizable as a simple quantum system, for example the two basis states may correspond to the horizontal and vertical polarization of a photon.
Superposition allows quantum computing systems to potentially achieve exponential speedups over their classical counterparts, due to the parallel computations on the probabilistic combinations of states.

Entanglement refers to the phenomenon by which qubits exhibit correlation with one another. 
In general, a set of $n$ entangled qubits exist as a superposition of $2^n$ basis states. 
Observing one or more qubits among them causes a collapse of their states, and alters the original superposition to account for the observed values of the qubits. 
For example, consider the 2-qubit system in the following initial state:
$$
\ket{\psi} = \frac{1}{\sqrt{3}}\ket{00} + \frac{1}{\sqrt{3}}\ket{01} + \frac{1}{\sqrt{6}}\ket{10} + \frac{1}{\sqrt{6}}\ket{11}
$$
Suppose a measurement of the first qubit yields a value of 0 (which can occur with probability $\frac{2}{3}$). Then, $\psi$ collapses into:
$$
\ket{\psi'} = \frac{1}{\sqrt{2}}\ket{00} + \frac{1}{\sqrt{2}}\ket{01}
$$
Note that the relative probabilities of the possible states are conserved, after accounting for the state collapse of the observed qubits. 

\paragraph{Quantum operators}
In classical computing, two fundamental logic gates (AND and OR) perform irreversible computations, i.e. the original inputs cannot be recovered from the output. 
Quantum gates (which operate on qubits) are constrained to be \textit{reversible}, and operate on the input state to yield an output of the same dimension.  
In general, quantum gates are represented by unitary matrices, which are square matrices whose inverse is their complex conjugate. 

An $n$-qubit system exists as a superposition of $2^n$ basis states. 
Its state can be described by a $2^n$ dimensional vector containing the coefficients corresponding to each basis state. 
For example, the $\ket{\psi}$ vector above may be described by the vector $[\frac{1}{\sqrt{3}}, \frac{1}{\sqrt{3}}, \frac{1}{\sqrt{6}}, \frac{1}{\sqrt{6}}]^T$ using the basis vectors.

Thus, a $n$-qubit quantum gate $H$ represents a $2^n \times 2^n$ unitary matrix that acts on the state vector. 
Two common quantum gates are the Hadamard and CNOT gates. The Hadamard gate acts on 1-qubit and maps the basis states $\ket{0}$ and $\ket{1}$ to $\frac{\ket{0}+\ket{1}}{\sqrt{2}}$ and $\frac{\ket{0}-\ket{1}}{\sqrt{2}}$ respectively. The CNOT gate acts on 2-qubits and maps $\ket{a,b}$ to $\ket{a,a \oplus b}$. In other words, the first bit is copied, and the second bit is flipped if the first bit is 1. 
The unitary matrices corresponding to the Hadamard and CNOT gates are: 
$$
H =  \frac{1}{\sqrt{2}} \begin{bmatrix}
1 & 1 \\
1 & -1 
\end{bmatrix}, 
\text{on the basis } [\ \ket{0}, \ket{1} \ ]
$$
$$
CNOT =   \begin{bmatrix}
1 & 0 & 0 & 0 \\
0 & 1 & 0 & 0 \\
0 & 0 & 0 & 1 \\
0 & 0 & 1 & 0 \\
\end{bmatrix}, 
\text{on the basis }[\ \ket{00}, \ket{01}, \ket{10}, \ket{11} \ ]
$$

The Pauli matrices ($\{\sigma_x,\sigma_y,\sigma_z\}$) are a set of three $2\times 2$ complex matrices which form a basis for the real vector space of 2 × 2 Hermitian matrices along with the $2\times 2$ identity matrix.

% The corresponding Pauli operators denote the observable corresponding to spin along the $k^{th}$ coordinate axis in the three-dimensional Euclidean space $\mathcal{R}^3$.
$$
\sigma_x = \begin{bmatrix}
0 & 1 \\
1 & 0 \\
\end{bmatrix},
\sigma_y = \begin{bmatrix}
0 & -i \\
i & 0 \\
\end{bmatrix},
\sigma_z = \begin{bmatrix}
1 & 0 \\
0 & -1 \\
\end{bmatrix}
$$

For a $d$ dimensional function space, the density operator $\rho$ represents a mixed state and is defined as:

$$
\rho= \sum_{i=0}^{2^d} p_i \ket{\psi_i}\bra{\psi_i}$$

where $\{\psi_i\}$ represent the computational bases of the $\mathcal{H}^{2^n}$ Hilbert space, the coefficients $p_i$ are non-negative probabilities and add up to $1$, and $\ket{\psi}\bra{\psi}$ is an outer product written in bra-ket notation. The expected value of a measurement $X$ can be obtained using the density operator using the following formula:

$$
\langle X \rangle  = \sum_{i} p_i \bra{\psi_i}X\ket{\psi_i} = \sum_{i} p_i \ tr(\ket{\psi_i}\bra{\psi_i}X)= tr(\rho X)
$$

where $tr$ denotes the trace of the matrix.

\section{Quantum Neural Network}
\label{sec:qnn}
Multiple research works~\cite{GUPTA2001355, NIPS2003_2363, journals/qip/SchuldSP14, Wan_2017,cao2017quantum,Daskin_2018, Farhi2018ClassificationWQ, shao2018quantum,Beer2020quantum} have proposed formulations for a quantum neural network(QNN) as a quantum analogue to a perceptron. \citet{NIPS2003_2363} were one of the earliest to propose a QNN which was modelled using a quantum circuit gate whose weights were learned using quantum search and piecewise weight learning. Several of these papers share a high level idea with respect to formulating the QNN through reversible unitary transforms on the data and then learning them through an approach analogous to the backpropagation algorithm. In this section, we present an overview of a QNN by breaking its components for learning a regression/classification problem in the quantum setting.

\subsection{Representing the input}
\label{ssec:qnn_input}
Inherently, the classical neural network computations are irreversible, implying a unidirectional computation of the output given the input. When mathematically posed, a classical NN computes the output $y$ from the input: $(x_1, x_2, \dots, x_d) \rightarrow y$. In contrast, quantum mechanics inherently depends on reversible transforms and a quantum counterpart for transforming the inputs to outputs for a NN can be posed by adding an ancillary bit to the input to obtain the output: $(x_1, x_2, \dots, x_d, 0) \rightarrow (x_1', x_2', \dots, x_d', y)$. \citet{Muthukrishnan99classicaland} show that such an operation can be always represented through a permutation matrix. To make the input representation unitary, we represent the input component of the vector $(x_1,\dots, x_d)$ through a quantum state $\ket{\psi}_{1,\dots,d}$. An ancillary dummy qubit can be added to $\ket{\psi}_{1,\dots,d}$ corresponding to the output $y$. The reversible transformation is thus rendered unitary in the quantum setting as: $\ket{\psi}_{1,\dots,d}\ket{0} \rightarrow \ket{\psi'}_{1,\dots,d}\ket{y}$ where $\ket{\psi'}_{1,\dots,d}$ represents the transformed input qubits. For multi-class classification problems, when the output labels cannot be captured by a single qubit, one can allocate $\mathcal{O}(logK)$ output qubits to represent the label where $K$ is the number of label classes.\\

QNNs can take as input purely quantum data or transformation of classical data into quantum states. When representing quantum data, $\ket{\psi}_{1,\dots,d}$ can be a superposition of the $2^{d}$ computational basis in the $d$-dimensional Hilbert space $\mathcal{H}^{2^{d}}=\mathcal{H}^{2} \otimes \dots \otimes \mathcal{H}^{2}$ where $\mathcal{H}^{2}$ represents the 2-dimensional Hilbert space with basis $\{\ket{0},\ket{1}\}$ and the basis for $\mathcal{H}^{2^d}$ are $\{\ket{0,0,\dots,0},\dots, \ket{1,1,\dots,1}\}$. Thus $\ket{\psi}_{1,\dots,d}$ can be denoted as $\ket{\psi}_{1,\dots,d} = \sum_{i=1}^{2^d}a_i \ket{z_i}$ where $a_i, i \in \{1,\dots,2^d\}$ represents the complex amplitudes assigned to computational basis states $\ket{z_i} \in \mathcal{H}^{2^{d}}$.

While exploiting truly quantum data is the eventual goal of developing QNN models, the majority of related works shift their focus to the immediate benefits derived from QNNs over classical data. To transform classical data to a quantum state represented through qubits, several popular strategies have been put to use. 
An easy strategy, popularly used by several QNN proposals \cite{Farhi2018ClassificationWQ}, is to binarize each individual component $x_i, i \in \{1,\dots,d\}$ of the input $x=(x_1, x_2, \dots, x_d)$ through a threshold, and then represent each binarized dimension $x_i$ as a corresponding $\ket{0}/\ket{1}$ qubit resulting in $x$ being represented as a computational basis in the $\mathcal{H}^{2^d}$ Hilbert space. This approach leads to a high loss of information contained in the data. 
To counter this, \citet{journals/corr/abs-1812-03089} suggest capturing a more fine-grained representation of $x$ as a superposition of computational basis in the $\mathcal{H}^{2^d}$ Hilbert space. 
For example, let $\ket{i}$ denote the computational basis  corresponding to the quantum state $\ket{0,\dots,1, \dots,0}$ with the qubit 1 in the $i^{th}$ position for each dimension $i \in \{1,\dots,d\}$. Then $x$ can be represented as a quantum state $\ket{\psi}_{1,\dots,d} = \sum_{i=1}^{d} b_i \ket{i}$ where $b_i = \frac{x_i}{||x||}$.\\

% $\ket{\psi}_{1,\dots,d} = 0.05 \ket{1} + 0.35 \ket{2} + 0.45 \ket{3} + 0.05 \ket{4} + 0.1 \ket{5} + 0.4 \ket{6} $

In parallel work, some strategies have been proposed in the continuous-variable architecture \cite{journals/corr/abs-1806-06871}, which encodes the input to quantum states through continuous degrees of freedom such as the amplitudes of the electromagnetic fields.
This approach avoids the information loss due to the discretization of continuous inputs, however at the cost of complexity of practical realization. 

\begin{figure}
    \centering
    \includegraphics[width=5in]{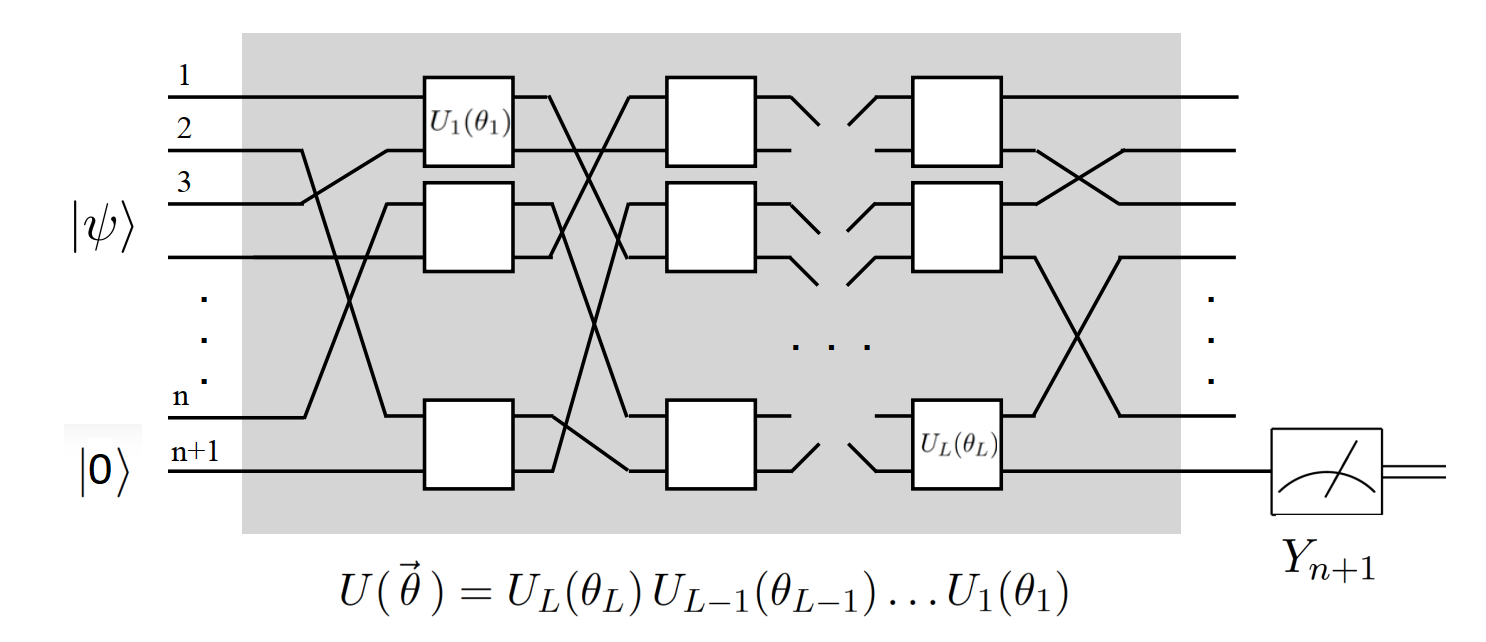}
    \caption{Illustration of a $L$ layered quantum circuit QNN where the input is $\ket{\psi}\ket{0}$ and the final output is measured through a Pauli-y operator on the ancillary bit (Figure from \cite{Farhi2018ClassificationWQ})}
    \label{fig:qnn}
\end{figure}

\subsection{Modeling the Quantum Network}
\label{ssec:qnn_modeling}
The quantum network has been most popularly modelled through learnable variational quantum circuits~\cite{Torrontegui2018} . 
A permutation matrix can be used to transform $\ket{\psi}_{1,\dots,d}\ket{0} \rightarrow \ket{\psi'}_{1,\dots,d}\ket{y}$ and therefore is the simplest technique for the QNN model. Mathematically, a square matrix $P$ is a permutation matrix if $PP^T =I$ and all entries of $P$ are either 0 or 1. However, the total number of distinct permutation matrices is a discrete set of size  $n!$ and therefore restricts the richness of representations that they can capture.
This transformation can be modelled more richly using unitary matrices, which are characterized by learnable free parameters. Any unitary matrix $U$ can be expressed as $U = e^{iH}$, where $H$ is a Hermitian matrix. Since every Hermitian matrix can be written as linear combinations of
tensor products of the Pauli matrices ($\{\sigma_x,\sigma_y,\sigma_z\}$) and the identity matrix($I$), the unitary matrix $U$ over $K$ bits can be written as 

$$U =exp\bigg[i \big(\sum_{j_1=0,\dots,j_K=0}^{3,\dots,3}\alpha_{j_1,j_2,\dots,j_K} \times (\sigma_{j_1} \otimes \dots \otimes \sigma_{j_K}) \big)\bigg]$$ where $\sigma_i$ denotes $\{I_{2\times2}, \sigma_x,\sigma_y,\sigma_z\}$ respectively for $\{i=0,1,2,3\}$ and $\alpha_{j_1,j_2,\dots,j_K}$ is the trainable free parameter. For notational brevity, we will denote a K bit unitary $U$ as $U^{K}(\Theta)$ where $\Theta$ is the set of all free parameters $\{\alpha_{j_1,j_2,\dots,j_K}\}_{j_1=0,\dots,j_K=0}^{j_1=3,\dots,j_K=3}$. For our input representation $\ket{\psi}_{1,\dots,d}\ket{0}$ we need a $d+1$ bit unitary matrix to transform this to the output $\ket{\psi'}_{1,\dots,d}\ket{y}$. Thus the simple variant of a quantum neural network, analogous to a single perceptron in the classical setting, uses a single unitary matrix of dimension $d+1$ and can be denoted by 
$$ U^{d+1}(\Theta)\ket{\psi}_{1,\dots,d}\ket{0}$$

% \begin{figure}[t]
%     \centering
%     \includegraphics[width=3in]{images/CV_analogues.png}
%     \caption{A table showing the analogues  (from \cite{Farhi2018ClassificationWQ})}
%     \label{fig:cv_analogues}
% \end{figure}

To capture detailed patterns in the input, a quantum neural network may be a cascade of several variational circuits, similar to a classical deep neural network. A sequential cascade of $L$ unitary matrices may be denoted as the following (we skip writing the $U^{d+1}_{i}$ for notational brevity):
$$\mathcal{U}(\Theta) = U_L(\Theta_L)U_{L-1}(\Theta_{L-1})\cdots U_1(\Theta_1)$$
where $U_{i}(\Theta_i)$ denotes the unitary matrix corresponding to the $i^{th}$ layer and $\Theta=\{\Theta_1,\dots,\Theta_L\}$ is the set of all parameters.\\

Some recent works~\cite{Beer2020quantum} have further increased the modeling complexity of $\mathcal{U}$ through a more direct inspiration from classical NNs: having multiple hidden units for every layer in the model. 
We introduce an additional notation for the mixed state density corresponding to the input state $\ket{\psi}_{1,\dots,d}$ as $\rho^{in}= \sum_{i=0}^{2^d} p_i \ket{\psi_i}\bra{\psi_i}$, where $\psi_i$ denote the computational basis of the $\mathcal{H}^{2^d}$ Hilbert space.
In~\cite{Beer2020quantum}, the first layer $U_{1}$ initializes a state of $\ket{0,\dots,0}$ of dimension $h$ (hidden state dimension) in addition to the input state $\ket{\psi}_{1\dots d}\ket{0}$. $U_{1}$ is applied to $\ket{\psi}_{1,\dots,d} \otimes \ket{0,\dots,0}_{h} \bra{0}$, where $\bra{0}$ corresponds to the ancillary output qubit. 
Here $U_{1}$ can be denoted as a sequential product of multiple unitary matrices $U_{1}= U_{1}^{m_1}\dots U_{1}^{2}U_{1}^{1}$ corresponding to $m_1$ number of perceptrons in layer $1$. 
This transformation is denoted as $X^1 = U^{1} (\rho_{in} \otimes \ket{0,\dots,0}_{h} \bra{0}){U^{1}}^{\dagger}$. 
From $X^1$, the density operator corresponding to the $h$ hidden state qubits and the output ancillary qubit are extracted using a partial trace operator, and fed to the next layer where the transforms are applied in the same way.
Having $m_l$ number of perceptrons in layer $l$ allows a greater degree of freedom to the QNN to capture patterns in the data.

In the continuous variable architecture, \citet{journals/corr/abs-1806-06871} model a QNN as a variational quantum circuit, with gaussian and non-gaussian gates used to implement linear and non-linear transformations. 

\subsection{Observing the Output}
\label{ssec:qnn_output}
\citet{journals/qip/SchuldSP14} describe several works~\cite{Menneer1995,Zak1998} where measuring the output from the network corresponds to the collapse of the superposition of quantum states to a single value, forming a close analogue to the non-linearity imposed in classical NNs through activation functions. \\

When the data is truly quantum in nature, the output state $\ket{y}$ corresponding to the input state $\ket{\psi}_{1,\dots,d}$ can be a pure computational basis or a mixed quantum state. Let the the mixed state density for the output state obtained from the QNN be denoted by $\rho^{out}$, corresponding to the last qubit in the final quantum state obtained after the unitary matrix operations. A popular measure of closeness between the observed and actual output quantum state is their fidelity, which when averaged over the training data can be mathematically represented as:

$$C = \frac{1}{N} \sum_{x=1}^{N} \bra{y_{x}}\rho^{out}_{x}\ket{y_{x}}$$

\citet{Beer2020quantum} show that the fidelity is a direct generalization of the classical empirical risk. When the the output state $\ket{y}$ for the input is mixed quantum state and not a computational basis, the fidelity expression can simply be modified to account for the case when $\ket{y}$ is mixed.

When the input data was originally in a classical form and the output is a classical scalar/vector value, measurement of the output state from the QNN has been the popular approach~\cite{Farhi2018ClassificationWQ,Wan_2017} to compute the cost function ($C$). \citet{Farhi2018ClassificationWQ} measure a Pauli operator, say $\sigma_y$ on the readout bit and denote this measurement by $Y$. Measuring $Y$ is probabilistic in the different possible outcomes, and hence an average of $Y$ is measured for multiple copies of the input $\ket{\psi}_{1,\dots,d}\ket{0}$. Averaging $Y$ computes the following: $$y^{out}=\bra{\psi_{1,\dots,d}\ 0}\mathcal{U}(\Theta)^{\dagger} Y \mathcal{U}(\Theta)  \ket{\psi_{1,\dots,d}\ 0}$$
The loss $C$ can now be defined as a mean squared error or $0/1$ loss with respect to this averaged value of $Y$ as:

$$C_{MSE}= \frac{1}{N} \sum_{x=1}^{N} |y_{x}-y^{out}_{x}|^{2} \ \  \text{or} \ \ C_{0/1}= \frac{1}{N} \sum_{x=1}^{N} \mathds{1}[y_{x}=y^{out}_{x}]$$

where $y_{x},y^{out}_{x}$ corresponds to the original output and averaged QNN output for input $x$.

\subsection{Learning network parameters}
\label{ssec:learning}
Similar to classical deep learning, the QNN parameters, $\Theta$ for $\mathcal{U}$, are learnt by using first-order optimization techniques to minimize a loss function over the dataset. The simplest gradient based update rule is the following:
$$
\Theta \ \leftarrow \Theta - \eta \frac{\partial C(\Theta)}{\partial \Theta}
$$
where $\Theta$ are the parameters being learnt, $C$ is the loss computed over the data and $\eta$ is the step-size. A second order estimate of the derivative of a function can be found using the finite difference method as:

$$\frac{dC(\Theta_i)}{d\Theta_i}= \frac{C(\Theta_i+\epsilon)-C(\Theta_i-\epsilon)}{2\epsilon}+ \mathcal{O}(\epsilon^2)$$

For this, the loss function $C$ for a particular value of the parameter set $\Theta_i$ for the unitary matrix $\mathcal{U}$ of layer $i$, needs to be estimated to within  $\mathcal{O}(\epsilon^3)$ and \citet{Farhi2018ClassificationWQ} show that this requires $\mathcal{O}(\frac{1}{\epsilon^6})$ measurements. This needs to be done for every layer parameter $\Theta_i$ independently resulting in $L$ such repetitions for a $L$-layer QNN.\\

Under a special condition on the unitary matrices $U(\Theta)$ for the QNN where they can be represented as $e^{i\Theta \Sigma}$ ($\Sigma$ being a tensor product of Pauli operators $\{\sigma_x,\sigma_y,\sigma_z\}$ acting on a few qubits), an explicit gradient descent update rule can be obtained. The gradient of the cost function $C(\Theta)$ with respect to the $\Theta_i$ for the $i^{th}$ layer parameters is given by:

$$ \frac{dC(\Theta)}{d\Theta_{i}}= 2 Im\big(\bra{\psi_{1,\dots,d}\ 0} U_{1}^{\dagger}\dots U_{L}^{\dagger}YU_{L}\dots U_{i+1}\Sigma_{i}U_{i}\dots U_{1} \ket{\psi_{1,\dots,d}\ 0} \big)$$

where $\Sigma_{i}$ is the tensor product of Pauli
operators corresponding to layer $i$ defined above and $Im()$ refers to the imaginary part. 
\citet{Farhi2018ClassificationWQ} make the interesting observation that $U_{1}^{\dagger}\dots U_{L}^{\dagger}YU_{L}\dots U_{i+1}\Sigma_{i}U_{i}\dots U_{1}$ is a unitary operation and can therefore be viewed as a quantum circuit of $2L+2$ unitaries each acting on a few qubits, therefore enabling efficient gradient computations. 
% and hence can be conditioned on the input state.

\subsection{QNN Variants}
\label{ssec:qnn_variants}
There have been multiple ideas proposed similar to a learnable QNN as described above. \citet{mitarai2018quantum} pose a problem through the lens of learning a quantum circuit, very similar to the QNN, and use a gradient-based optimization to learn the parameters. \citet{Romero_2017} introduce a quantum auto-encoder for the task of compressing quantum states which is optimized through classical algorithms. \citet{ngoc2020tunable} propose an alternate QNN architecture only using multi-controlled NOT gates and avoiding using measurements to capture the non-linear activation functions of classical NNs. \citet{zhao2019qdnn} suggest interleaved quantum structured layers with classical non-linear activations to model a variant of the QNN. Multiple ideas~\cite{mitarai2018quantum,zhao2019qdnn,journals/corr/abs-1812-03089} utilise a hybrid quantum-classical approach where the computation is split so as to be easily computable on classical computers and quantum devices.

\subsection{Practical implementations of QNNs}
\label{ssec:qnn_practical}

While modelling a QNN has been a hot topic in the field of quantum deep learning, several of the algorithms cannot be practically implemented due to the poor representation capability of current quantum computing devices. There has been considerable research in the field of practically implementing QNNs~\cite{behrman2002quantum} and developing hybrid quantum-classical algorithms which can successfully perform computations using a small QRAM. 

Early works in practically implementing QNNs used the idea of representing the qubits through polarized optical modes and weights by optical beam splitters and phase shifters~\cite{altaisky2001quantum}. Parallely, \citet{Behrman2000} proposed implementing the QNN through a quantum dot molecule interacting with phonons of a surrounding lattice and an external field. Such a model using quantum dotshas been extensively studied  since~\cite{Toth_1996,831067,Altaisky2014}.

\citet{Korkmaz_2019} used a central spin model as a practical implementation of a QNN using a system of 2 coupled nodes with independent spin baths. A similar idea was earlier proposed by \citet{Deniz2017} using a collisional spin model for representing the QNN thereby enabling them to analyse the Markovian and non-Markovian dynamics of the system.

The majority of the recent research in the area of practical implementations of QNNs has been centered on simulating quantum circuits on Noisy Intermediate-Scale Quantum Computing (NISQ) devices. \citet{Juncheng2015} presented a neuromorphic hardware co-processor called Darwin Neural Processing Unit (NPU) which is a practical implementation of the Spiking Neural Network (SNN)~\cite{Tavanaei_2019,NIPS2018_7417}, a type of biologically-inspired NN which has been popularly studied recently.

\citet{potok2017study} conduct a study of performance of deep learning architectures on 3 different computing platforms: quantum (a D-Wave processor~\cite{Johnson2011Quantum}), high performance, and neuromorphic and show the individual benefits of each. \citet{Tacchino2019} experimentally use a NISQ quantum processor and test a QNN with a small number of qubits. They propose a hybrid quantum classical update algorithm for the network parameters which is also parallely suggested by \cite{tacchino2019quantum}.

\section{Complex Quantum Neural Network Architectures}
\label{sec:quantum_cnn_rnn}

\subsection{Quantum CNNs}
\begin{figure}
    \centering
    \includegraphics[width=3in]{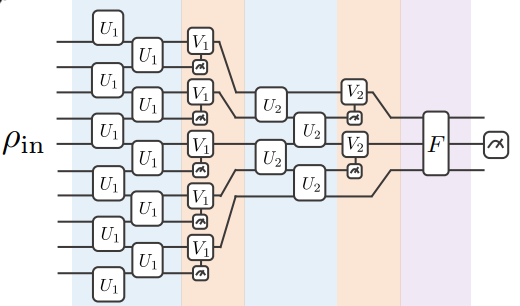}
    \caption{The Quantum CNN architecture proposed by \citet{Cong_2019}}
    \label{fig:quantum_cnn}
\end{figure}

\citet{Cong_2019} propose a quantum CNN through a quantum circuit model adapting the ideas of convolutional and pooling layers from classical CNNs. 
The proposed architecture (shown in Figure \ref{fig:quantum_cnn}) is similarly layered, however it differs in the fact that it applies 1D convolutions to the input quantum state (contrary to 2D/3D convolutions on images). 
The convolutional layer is modeled as a quasi-local unitary operation on the input state density $\rho_{in}$. 
This unitary operator is denoted by $U_i$ and is applied on several successive sets of input qubits, up to a predefined depth.
The pooling layer is implemented by performing measurements on some of the qubits and applying unitary rotations $V_i$ to the nearby qubits. 
The rotation operation is determined by the observations on the qubits. 
This combines the functionality of dimensionality reduction (the output of $V_i$ is of lower dimension) as well as non-linearity (due to the partial measurement of qubits). 
After the required number of blocks of convolutional and pooling unitaries, the unitary $F$ implements the fully connected layer.  
A final measurement on the output of $F$ yields the network output. 

Similar to classical CNNs, the overall architecture of the quantum CNN is user-defined, whereas the parameters of the unitaries are learned. 
The parameters are optimized by minimizing a loss function, for example by using gradient descent using the finite difference method described in Section~\ref{ssec:learning}. 
\citet{Cong_2019} demonstrate the effectiveness of the proposed architecture on two classes of problems, quantum phase recognition (QPR) and quantum error correction (QEC).

More recently, \citet{Kerenidis2020Quantum} identify the relation between convolutions and matrix multiplications, and propose the first quantum algorithm to compute the forward pass of a CNN as a convolutional product. 
They also provide a quantum back propagation algorithm to learn network parameters through gradient descent. 
In an application of CNNs, \citet{journals/pr/ZhangCWBH19} and \citet{journals/corr/abs-1901-10632} propose special convolutional neural networks for extracting features from graphs, to identify graphs that exhibit quantum advantage. 

\subsection{Hybrid CNNs}
\citet{henderson2019quanvolutional} introduce the \textit{quanvolutional} layer, a transformation based on random quantum circuits, as an additional component in a classical CNN, thus forming a hybrid model architecture. 
Quanvolutional layers consist of multiple quantum filters, each of which takes a matrix of 2D values as input, and outputs a single scalar value.
Similar to convolutional filters, the operations are iteratively applied to subsections of the input. 
Each quantum filter is characterized by an encoder, random circuit, and decoder, where the encoder converts the raw input data into an initialization state to be fed into the random circuit and the output from the circuit is fed to the decoder which yielding a scalar value. \cite{henderson2019quanvolutional} do not present a learning methodology to optimize the random circuits since the quanvolutional layer has no learnable parameters.
However, the experimental results suggest that the quanvolutional layer performed identically to a classical random feature extractor, thus questioning its utility. 

\subsection{Quantum RNNs}

There has also been several interesting suggestions to the front of developing quantum variants of recurrent neural networks. \citet{hibatallah2020recurrent} propose a quantum variant of recurrent neural networks(RNNs) using variational wave-functions to learn the approximate ground state of a quantum Hamiltonian. \citet{roth2020iterative} propose an iterative retraining approach using RNNs for simulating bulk quantum systems via mapping translations of lattice vectors to the RNN time index. 
Hopfield Networks~\cite{hopfield-neural-networks-and-1982} were a popular early form of a recurrent NN 
% based on underlying Lyapunov functions.
for which several works \cite{Rebentrost_2018,Tang_2019,Rotondo_2018} have proposed quantum variants.

\section{Quantum inspired Classical Deep Learning}
\label{sec:applications}

% \subsection{Applications to Natual Language Processing}
Quantum computing methods have been applied to classical deep learning techniques by several researchers. 
\citet{journals/corr/AdachiH15} suggest a quantum sampling-based approach for generative training of Restricted Boltzmann Machines, which is shown to be much faster than Gibbs sampling. 
\citet{C6SC05720A} use quantum mechanical (QM) DFT methods to train deep neural networks to build an molecular energy estimating engine. 
\citet{journals/ijon/LiXCJ19} propose to use quantum based particle swarm optimization to find optimal CNN model architectures. 
\citet{da_Silva_2017} propose a quantum algorithm to evaluate the performance of neural network architectures. 
\citet{Behera2004} use a quantum RNN variant to simulate a brain model, and use it to explain eye tracking movements. 

\paragraph{Natual Language Processing}
\citet{Clark2008ACD, coecke2010mathematical} introduce a tensor product composition model(CSC) to incorporate grammatical structure into algorithms that compute meaning.  \citet{Zeng_2016} show the shortcomings of the CSC model with respect to computational overhead and resolve it  using QRAM based quantum algorithm for the closest vector problem.

% $$
% L_{DN\_soft}(x,y') = L_{CE}(\hat y', y') + \beta \cdot (1-P_N(x))\cdot L_{CE}(\hat y, y')
% $$
% \vspace{5pt}
% $$
% L_{DN\_hard}(x,y') = L_{CE}(\hat y', y') + \beta \cdot (1-N(x))\cdot L_{CE}(\hat y, y')
% $$
 
\citet{10.1145/2484028.2484098,Zhang2018EndtoEndQL} suggest a language modelling approach inspired from the quantum probability theory which generalizes \cite{Sordoni_2014}. \citet{Zhang_2018} present an improved variant of the quantum language model which has higher representation capacity and can be easily integrated with neural networks.

\citet{Galofaro_2018} tackle the problem of  typification of semantic relations between keyword couples in hate and non-hate speech using quantum geometry and correlation. \citet{Li_2018} utilise the Hilbert space quantum representation by assigning a complex number relative phase to every word and use this to learn embeddings for text classification tasks. \citet{oriordan2020hybrid} recently present a hybrid workflow toolkit for NLP tasks where the classical corpus is encoded, processed, and decoded using a quantum circuit model.

\section{Conclusion}
\label{sec:conclusion}

Quantum computing and deep learning are two of the most popular fields of research today. 
In this work, we have presented a comprehensive and easy to follow survey of the field of quantum deep learning.
We have summarized different schemes proposed to model quantum neural networks (QNNs), variants like quantum convolutional networks (QCNNs) and the recent progress in quantum inspired classic deep learning algorithms.
There is a tremendous potential for collaborative research at the intersection of the two fields, by applying concepts from one to solve problems in the other. 
For example, \citet{Levine_2019} demonstrate the entanglement capacity of deep networks, and therefore suggest their utility for studying quantum many-body physics.

\bibliographystyle{plainnat}
\bibliography{references.bib}

\end{document}